\documentstyle[twocolumn,aps,prl,overcite]{revtex}
\input epsf

\tighten
\begin{document}


\title{Limitations in Using 
Luminosity Distance to  Determine the Equation of state of the
Universe
 }

\author{Irit Maor$^{(1)}$, Ram Brustein$^{(1)}$,
and Paul J. Steinhardt$^{(2)}$
}

\address{$(1)$Department of Physics, Ben Gurion University,
Beer Sheva
84105, Israel \\
$(2)$ Department of Physics, Princeton University, Princeton, NJ
08540 USA
}

\maketitle


\begin{abstract}
Supernova searches 
have been been
suggested as a  method  for determining  precisely 
the current value  and time variation of 
the equation of state, $w$,  of the  dark energy component 
responsible for the accelerated expansion of the  Universe.
We show that the method
is fundamentally limited by the fact that  luminosity distance 
depends on  $w$  through a multiple integral relation
that smears out information about $w$ and its time variation.
The effect  degrades
the resolution of $w$ that can be obtained from   current data. 
\end{abstract}

Recent observations suggest that most of the energy density of the
Universe consists of a dark energy component with negative pressure
that
causes the expansion rate of the Universe  to
accelerate.\cite{Bahcall}
A key challenge for cosmology and for fundamental physics is to 
determine the nature of the dark energy.  One possibility is that
the 
dark energy consists of vacuum energy or cosmological constant, in
which case
the equation of state is  $w\equiv p/\rho =-1$, where $p$ is the 
pressure and $\rho$ is the energy density of the dark energy.
An alternative  is quintessence,\cite{Caldwell} a time-evolving,
spatially
inhomogeneous energy component with negative pressure, such as a 
scalar field slowly rolling down a potential.   
For quintessence, the equation of state is typically a function of 
 redshift, $w(z)$, whose   value   differs from -1.  
Hence, a precise measurement of $w$ today and its time variation 
could distinguish between the two possibilities and provide
important
clues about the dynamical properties of dark energy.

Searches for type Ia supernovae at deep  redshift have 
provided the most direct evidence that the expansion rate of the
Universe is accelerating.\cite{SCP,HZS}
The supernovae appear to be standard candles
which can be used to measure the luminosity distance-redshift
relation.  By measuring 50  supernovae out to  redshift near $z=1$, 
the Supernova Cosmology Project (SCP)\cite{SCP}
and the High-$z$  Survey\cite{HZS}  Project have
each found strong evidence that the Universe is accelerating and
that
the equation of state of the dark energy component is
negative.\cite{EOS}

\begin{figure}
 \epsfxsize=3.1 in \epsfbox{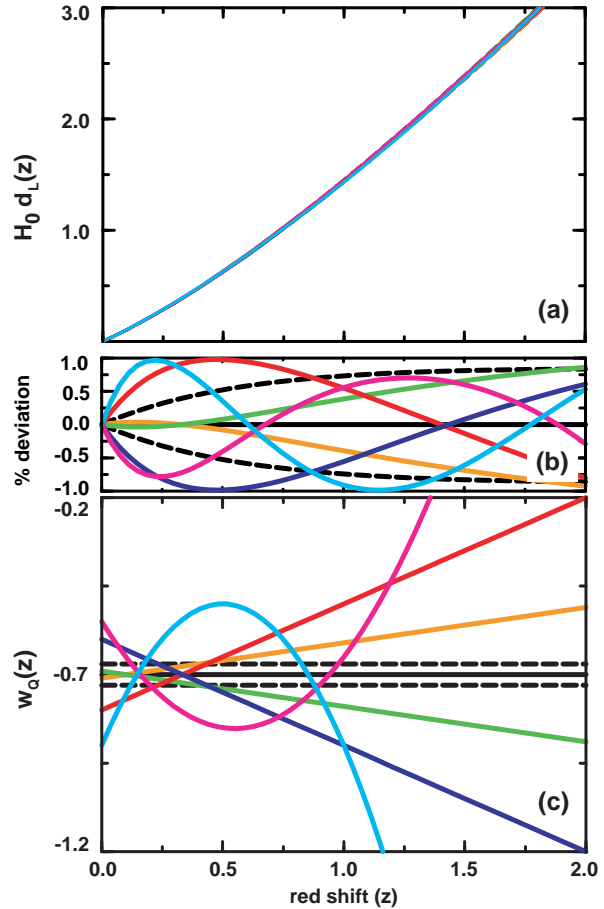}
 \caption{ (a) The luminosity distance $H_0 d_L(z)$ for nine choices
 of equation of state $w_Q(z)$ for the dark energy shown in (c),
where
 $H_0$ is the current value of the Hubble parameter.
 All models have $\Omega_m=0.3$.
 (b) Illustrates that the percentage deviation of $d_L(z)$ from a 
 a cosmological
 model with $\Omega_m=0.3$, $\Omega_Q=0.7$ and  $w_Q=-0.7 = const.$
 is less than 1\%.  If one artificially restricts
 $w_Q$ to be constant, then the range of models 
  collapses to the 
 region between the dashed lines.
 }
 \end{figure}

A supernova search  extended to greater $z$ 
can make a much more  precise
determination of the luminosity distance as a function of red 
shift,\cite{SNAP,Hut} 
$d_L(z)$,
perhaps to better than 1\% uncertainty out to  redshift $z=2$.
(1\% is probably an optimistic estimate of the limiting
systematic 
uncertainty.)
Does this enable a precise determination of the equation of state 
of the dark energy component and its time variation?   
As we show in this paper, the answer is no.  The inherent limitation
is
theoretical: the luminosity distance depends on $w(z)$  through 
a multiple-integral relation that  smears out detailed information
about
$w(z)$.   Consequently, the value of $w(z)$ today  is poorly
resolved
and no  useful constraint  can be obtained about its time variation.

The problem can be immediately appreciated from Fig.~1, 
which compares $d_L(z)$ for an assumed cosmological model
[$\Omega_m=0.3$, $\Omega_Q=0.7$ and  $w_Q=-0.7 = const.$, where
$\Omega_{m,Q}$ is the ratio of the (matter, quintessence) energy
density to the critical density] 
with eight  other models chosen as examples where $d_L(z)$ 
is nearly degenerate with the assumed model.  In this figure
and throughout the paper, we assume the Universe is cosmologically
flat and the speed of light $c=1$.
(Henceforth, we use the subscript Q to label the dark energy
component,
be it quintessence or cosmological constant.)
Figure~1a shows that $d_L(z)$ is nearly identical for the set of
models
as individual curves can hardly  be distinguished.  Fig.~1b displays
the percentage deviation of $d_L(z)$  from the assumed model, where
it can be seen that the deviation is less than one per cent  out
to  redshift $z=2$.  Fig.~1c then shows 
$w_Q(z)$ for the respective models.   The striking result is the
wide
range of $w_Q(z)$ that produces nearly the same $d_L(z)$ as the 
assumed model. If one expands $w_Q(z) = w_0 + w_1 z + w_2 z^2 +
\ldots,$
then, for this particular collection of models,
$w_0$  varies between -0.55 and  -0.9 (a total span of 50\%
about the assumed value, $w_Q= -0.7$) and $w_1 = dw_Q/dz_0$ 
varies between -1.1 and +1.6.  (The subscript ``0" 
refers to  present-day values of parameters.)

Note that  the degenerate  models chosen for the illustration
span a larger range of 
$|dw_Q/dz| = {\cal O}(1)$   than most 
realistic models predict. Typically, $|dw_Q/d z_0| \ll 1$
because $w_Q(z)$  is bounded in most cases to lie between
-1 and +1 in order that the dark energy obey the positive energy
condition and be stable under perturbations.
The large uncertainty in $w_1=dw_Q/dz_0$ means that 
little  useful information is obtained about the magnitude or 
sign of the time variation of $w_Q$. 
Also, $w_0$ is poorly resolved.    The resolution
of $w_Q(z)$
degrades significantly further if one includes the  uncertainty 
in  $\Omega_m$, as shown in
 in Fig.~2 
(see discussion below). 

Our conclusion may seem at odds with 
some projections of  what
can be obtained in future supernova searches.\cite{SNAP,Hut}
Many analyses assume $w_Q=const.$
 If we impose this condition, then the range of models that
fit  collapses to the narrow region between the 
dashed lines in Fig.~1c, giving a misleading impression that 
$w_Q(z)$ is well-resolved. However, if consideration is extended to
models in which $w_Q$ is $z$-dependent, such as the
linear form  $w_Q(z) = w_0 + w_1 z$, 
 the result is dramatically different.
A very wide range of $(w_0, \, w_1)$  produces nearly identical
$d_L(z)$
because the differences are smoothed out by the multi integral
relation
derived below
between luminosity distance and $w_Q(z)$.  This degeneracy  accounts
for the results found in Figs.~1c and 2, 
but is  missed if one artificially restricts $w_Q$ to 
be constant. Notice in Fig.~1c that including  non linear forms for
$w_Q(z)$
enhances the uncertainty in $w_0$ and $w_1$ even further.
Among studies which have considered time varying  $w_Q(z)$,
our results agree with some\cite{Efst,Ratra,Alb}
but seem significantly less optimistic than others.\cite{Hut,Staro}
In the latter cases, the assumptions about the observations
are similar but various subtle factors, such as the 
use of fitting functions rather than exact expressions for $d_L(z)$
or imposing the constraint $w_Q >-1$, 
combine numerically to reduce  artificially the degeneracy.
Extending searches to yet deeper  redshift 
($z>2$) does not help either
because the effect of quintessence on
$d_L(z)$ is proportional to $\Omega_Q$ 
which  becomes very small
at deep  redshift. 

The key to understanding these conclusions is  the 
relation between luminosity distance and the equation of state.
The luminosity distance is related to the Robertson-Walker scale
factor
$a(t)$
through the equation 
\begin{equation} 
d_L(z) = (1+z) \int_a^{a_0} \frac{da'}{\dot{a'} a'}
= (1+z) \int^{1+z}_1 \frac{d x}{H}
\end{equation}
where 
the
 redshift $z$ satisfies
$1+z \equiv a_0/a $, and $H$ is the Hubble parameter, $H^2 = H_0^2
(\rho_T(z)/\rho_T(0))$.  We assume a flat Universe. 
The subscript ``T" refers to the total equation of state of the 
the combined matter-quintessence fluid.
Integrating the energy conservation equation,
\begin{equation} \label{cons}
\dot{\rho}_T  =   -3 H (1 + w_T) \rho_T 
\end{equation}
we can reexpress  $H^2  = H_0^2
[\rho_T(z)/\rho_T(0)]$ as
\begin{equation} \label{hint}
H^2 = H_0^2 \,  \frac{\rho_T(z)}{\rho_T(0)}=H_0^2 \,
{\rm exp}\, \left[3 \int_1^{1+z} \hspace{-.2in} d \, {\rm
ln}\, x \, \left(1+ w_T\right) \right]
\end{equation}
and the luminosity distance as
\begin{equation} \label{lum1}
d_L(z) = \frac{1+z}{H_0} \int\limits_1^{1+z}\hspace{-.05in}
 d x' \, {\rm exp}
 \left[ - \frac{3}{2} \int\limits_1^{x'}\hspace{-.05in} d \, {\rm
 ln}\, x \, \left(1+ w_T\right) \right].
 \end{equation}

Equation~(\ref{lum1}) shows that the luminosity distance depends on a 
double integral over the {\it total} equation of state, $w_T(z)$.  
One integral
is required to obtain the total luminosity distance from the present
back to  redshift $z$.  The integrand depends on $H$ which is itself
related to $w_T$ through the integral relation, Eq.~(\ref{hint}). 
To distinguish different forms of dark energy, though, we need to 
determine $w_Q(z)$, the equation of state of the dark energy
component.
Assuming that the Universe contains only pressureless matter
(baryonic and cold)
and dark energy, then $w_T = w_Q \Omega_Q$, where $\Omega_Q$  is
itself
related to $w_Q(z)$ through an integral relation. In particular,
using
the energy conservation analogous to Eq.~(\ref{cons})
for the dark energy component alone $(p_Q, \, \rho_Q)$, one obtains
\begin{equation}
\frac{\rho_Q(z)}{\rho_Q(0)} = 
 {\rm exp}\, \left[3 \int_1^{1+z} d \, {\rm
ln}\, x \, \left(1+ w_Q\right) \right];
\end{equation} 
combined with Eq.~(\ref{hint}), $w_T = w_Q \Omega_Q = w_Q
\rho_Q/\rho_T$
can be reexpressed as
\begin{equation} \label{lum2}
w_T(z) = w_Q(z)  \left\{ 1\!+\! \frac{\Omega_m}{\Omega_Q}
{\rm exp}\hspace{-.05in} \left[-3 \hspace{-.05in}\int\limits_1^{1+z}
\hspace{-.05in} d \, {\rm ln}\, x \, \left(w_Q\right) 
\right]\right\}^{-1}\hspace{-.15in},
\end{equation}
where $\Omega_{m,Q}$ refers to the current values.
Together with Eq.~(\ref{lum1}), this expression constitutes the 
integral relation between luminosity distance and $w_Q(z)$
that underlies the degeneracy problem.

To express the degeneracy problem  quantitatively, 
we have  found the maximum likelihood values 
of $w_0$ and $w_1$
based on current SCP supernova data,\cite{SCP}
 which has measured 50 supernova out to  redshift 
$z \approx 1$.  
For simplicity, we have assumed $w_Q(z) = w_0 + w_1 z$; including 
more general functions of $z$ only degrades the resolution further.
Furthermore, we have repeated the 
analysis based on simulated data from an idealized
experiment  which measures thousands of supernovae out to  redshift
$z = 2$. The simulated data
assumes a  cosmological model 
with $\Omega_m=0.3$, $\Omega_Q=0.7$ and $w_Q=-0.7= const.$
For the idealized experiment,
the absolute magnitude measurement for a single supernova
is taken to have   a  statistical variance of 0.15 and a 
systematic measurement  error of 0.02.
The supernovae 
are divided into 50 bins between $z=0$ and $z=2$
such that the statistical 
and systematic errors  averaged over a bin 
are comparable, resulting in an error in $d_L(z)$ for
a given bin equal to 1.4\%.  
We assume that other types of observations have constrained
$\Omega_m$ to 
lie between 0.2 and 0.4, say, and marginalize $\Omega_m$ over that 
range. 
For the  purposes of illustration, we assume that $w_Q(z)$ is a
linear
function of $z$  parameterized by $w_0$ and $w_1$.  As shown in 
Fig.~1c, including more general forms for $d_L(z)$ only worsens the
degeneracy problem.

We have considered  two ways of treating the systematic errors.
For Case I, we assume  that systematic  errors are random and 
uncorrelated from bin to bin and perform a likelihood analysis 
over the 50 bins with 1.4\% error each to determine the
uncertainty in $d_L(z)$. 
(To obtain 1.4\% per bin for 50 bins requires measuring thousands of
supernovae.)
In Case II, we assume there is  negligible statistical uncertainty
but 
correlated systematic error of 1\%.  
Examples of Case II  errors are those  due to calibration,
dust, or evolution of supernovae.
In this case, all models which
predict $d_L(z)$ within 1\% of the assumed cosmological 
model for all $z$ between
0 and 2 are 
deemed indistinguishable. As it turns out,
the 95\% confidence contours in Case I are roughly equivalent
to the
indistinguishability region of Case II, so both cases give
comparable 
results.

For the current data, likelihood  analyses based on the assumption
that 
$w_Q$ is constant have reported a resolution of $-1 < w_Q <-0.6$ at
the 95\% level.\cite{EOS}  When we repeat the analysis assuming 
a linear form for $w_Q(z)$ and making no  prior assumptions about 
$w_0$ and $w_1$, we find that neither parameter is well-determined.
The degeneracy obliterates the resolution of both quantities: 
$w_0$
can vary between -3.2 and -0.4 (95\% confidence)
and $w_1$  can vary between -11.8
and 11.0.
Notice the enormous  range of $w_0$; the 99\% confidence contour
includes positive values, 
so one cannot even be sure that the Universe is accelerating today.
In cases where the Universe is not accelerating today, we can 
still conclude that it must have been accelerating recently because 
$w_1$ is highly negative whenever  $w_0$ is positive.
One could argue that allowing large values of $w_1$ so that
$w_Q(z)$ becomes much less than -1  or greater than +1 
between $z=0$ and  $z=1$ (the range of current observations)
is unphysical based on the positivity and stability conditions
that apply to most (but not all) forms of dark energy.
Adding this theoretical constraint, the 95\%
range for $w_0$ is found 
to lie between -0.5 and -1.0 (95\% confidence), in which case 
 one may conclude 
that the Universe is accelerating today.  However, one should beware
that our estimates are optimistic in assuming that $w_Q(z)$ has only 
a linear
and constant term.  
A safer assessment would be that, assuming positivity and stability
but no other prior about $w_Q$, one can conclude
from present data that the Universe is accelerating today, but $w_Q$
is very poorly resolved, and $d w_Q/dz$ can range anywhere within
the 
imposed positivity and stability constraints.

For the idealized experiment,
the  likelihood contours 
span a substantial range of $(w_0, \, w_1)$, as
shown in Fig.~2. 
In the shaded ellipses the figure shows  likelihood region
if one assumes prior knowledge that $\Omega_m=0.3$ precisely.
The contours stretch along a curve in the 
$w_0-w_1$ plane  which corresponds to a near-degeneracy.
It is this degeneracy
that dashes hopes of using luminosity distance to measure both 
the current value and time derivative of $w_Q$.  Marginalizing over 
$\Omega_m$ expands the contours along a direction nearly orthogonal
to the degeneracy curve. See the large black contours in Fig.~2.
Within 
the 95\% confidence region, $w_Q$  spans a range equal to 
more than 35\% of the assumed value ($w_Q=-0.7$), 
and $w_1=d w/dz_0$ ranges 
between +0.3 and -1.1. 
Expanding the fit to include non linear $w_Q(z)$
would  expand the region even more (see Fig.~1).

Another approach for measuring the time evolution of $w_Q(z)$ is
object counts, where the objects might be
galaxies, clusters,  or halos:
$ \frac{ d N}{d\Omega} = n_c r^2 dr,$
 where $d N$ is the number of objects in a comoving volume
 element $r^2 d r d \Omega$ for coordinate distance $r$ and 
 solid angle $\Omega$.   One assumes that the number density of 
 objects per comoving volume, $n_c$, is constant or some known
function of
 $z$.
 The distance  $r$ is related to the 
 luminosity distance by
 $d_L(z)= (1+z) r$  (if we normalize 
 the Friedmann-Robertson-Walker scale
  to be unity today). Hence, we can write 
 \begin{equation}
 \frac{d N}{d z d \Omega} =
 n_c r^2\frac{ d r}{d z} = n_c
\frac{d_L^3}{(1+z)^4}\left[\frac{(1+z) d_L'}{d_L}
 -1\right]
 \end{equation}
 where prime represents derivative with respect to z.
 The novel feature here is $d_L'(z)$, which entails one less
integral
  of $w_Q$
 than $d_L$, and, hence, perhaps an  improved resolution.
 Newman and Davis\cite{Davis} suggest that near-future observations
of 
 the number of  dark matter halos as a function of their circular
velocity
 and  redshift 
can determine $\frac{d N}{d z d \Omega}$ 
to within a few percent.
Assuming a resolution of
2.5\% between $z=0.7$ and $z=2$ (more optimistic than their
example),
 we find
 combinations of $w_0$ and $w_1$ for which  $d N/dz d \Omega  $
 is indistinguishable from the 
 assumed cosmological model.
 The indistinguishability region  coincides approximately
 with the 95\% confidence contours in Fig.~2.  Hence, objects counts  
are subject  to essentially the same degeneracy problem as
supernova searches.

\begin{figure}
\epsfxsize=3.1 in \epsfbox{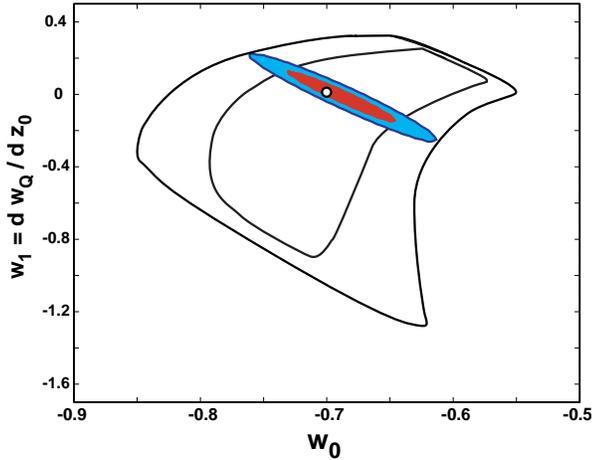}
  \caption{ 68\% and 95\% confidence contours in the 
  $(w_0, w_1)\equiv (w_Q(z=0), \, d w_Q/d z_0)$ plane
  for an idealized experiment
  which measures thousands of supernovae between $z=0$ and
    $z=2$.  The supernovae are  divided into 
  50 bins 
  with a net  error of 1.4\% per bin.
 The example assumes a  model with $\Omega_m=0.3$, 
  $\Omega_Q=0.7$, $w_Q=-0.7 = const.$, indicated by the 
  circle. The thin shaded ellipses are 68\% and 95\% confidence  
   contours if one assumes
  $\Omega_m$ is fixed to be precisely 0.3.  The broader black
contours 
  are the result if $\Omega_m$ is marginalized over the range 
  0.2 to 0.4.
           }
            \end{figure}

Our analysis has shown that the luminosity distance- redshift
relation and similar classical cosmological measures 
are limited in their ability to resolve
$w_Q(z)$, even under  optimistic 
assumptions (an absolutely flat Universe, rather stringent priors
for 
$\Omega_m$, exceptional accuracy in determining luminosity distance, 
{\it etc.})  
These conclusions hold unless the  errors can be reduced by 
at least 
1 order of magnitude or
some complementary experiment can break the degeneracy.  
If a method could be found to 
reduce  considerably the uncertainty in $\Omega_m$ --
in Fig. 2,  we  assumed $\Omega_m$ to be in the range 0.2 to 0.4 -- 
 the degeneracy region 
 shrinks along one direction. This would improve
 the resolution, but only modestly  
 because  there remains the second degeneracy
direction  shown in the figure.
It should be noted that reducing the uncertainty in $\Omega_m$ will 
be difficult. Most methods for determining $\Omega_m$ are 
dependent on some assumption about $w_Q$.  In the case of the 
CMB anisotropy, 
for example,  a degeneracy arises such that, for the same
high-precision
data, one can derive different values of $\Omega_m$ depending on 
what assumption is made about $w_Q(z)$.\cite{degen}
Of course,
if one assumes $|dw_Q/dz| \ll 1$ (or some other prior), then
deep supernova searches and galaxy counts can resolve
$w_Q(z=0)$ with impressive precision, but the value depends strongly
on the particular theoretical assumption.

Our conclusions also undermine claims that  the supernova 
and object count searches can 
determine the future fate of the  Universe.  Since 
the observations cannot distinguish whether
$dw_Q/d z_0$  is  positive or negative, they cannot distinguish 
whether $w_Q$ will remain  negative or become positive in the 
future, and, hence, whether the Universe will
accelerate ever faster 
or cease accelerating altogether.

We thank A. Albrecht, M. Davis, G. Efstathiou  and J. Newman
for helpful comments, D. Oaknin for valuable programming
assistance, and I. Wasserman for useful discussions and
suggestions on the paper. Additionally we wish to thank
E. Linder and A. Goobar for detailed comments on
the published version and for pointing out some errors
which prompted us to make this revision.
This research was supported by the US Department of Energy grant
 DE-FG02-91ER40671 (PJS).

\end{document}